\documentclass[twocolumn]{aastex63}
\usepackage{graphicx}
\usepackage{blindtext}
\usepackage{amsmath}
\usepackage{mathtools}
\usepackage{multirow}
\usepackage{hyperref}
\hypersetup{
	colorlinks	= true,
	linkcolor	= red,
	urlcolor	= cyan,
	citecolor	= blue
}

\newcommand{\iraclis}{\mbox{\textit{Iraclis}}}
\newcommand{\wayne}{\mbox{\textit{Wayne}}}
\newcommand{\taurex}{\mbox{$\mathcal{T}$-REx}}

\begin{document}

% TITLE
\title{Testing known and unknown systematics in HST/WFC3 spatial scans with the \textit{Wayne} simulator}

\correspondingauthor{A. Tsiaras}
\email{angelos.tsiaras.14@ucl.ac.uk}

\author{A. Tsiaras}
\affiliation{Department of Physics \& Astronomy, University College London, Gower Street, WC1E6BT London, United Kingdom}

\author{J. Ozden}
\affiliation{Department of Physics \& Astronomy, University College London, Gower Street, WC1E6BT London, United Kingdom}

\begin{abstract}
The Wide Field Camera 3  is one of the instruments currently onboard the  Hubble Space Telescope and, since 2012, with the use of the spatial scanning technique, has provided the largest number of observed exoplanetary atmosphere, ranging from super-Earths to hot Jupiters. This technique enables the observation of bright targets without saturating the sensitive detectors, but at the same time requires more complicated data reduction and calibration techniques to extract the planetary signal from the observations. In absence of absolute calibration sources, the validation of current data analysis techniques is not possible as the planetary signal is not known. Here, we demonstrate how simulated observations can help us understand the effect of different analysis processes, and potential unknown systematics on the final transmission spectra of exoplanets. We test and validate the robustness of two of the most precise WFC3 exoplanetary spectra -- HD\,209458\,b and 55\,Cancri\,e -- against three different known and potential sources of systematics. In addition, we identify the horizontal shifts seen in WFC3 observations as the most important source of systematic errors in the planetary spectra, concluding that a precision better that 1\% of a pixel is necessary.
\end{abstract}

% TITLE

% INTRODUCTION
\section{INTRODUCTION} \label{sec:introduction}

In the last two decades, transit spectroscopy has been the most valuable tool to characterise exoplanetary atmospheres. One of the instruments with the largest contribution to our current knowledge on atmospheric studies of exoplanets is the Hubble Space Telescope (HST). Since the first detection of an exoplanetary atmosphere with transmission spectroscopy \citep{Charbonneau2002}, HST data have revealed atomic, ionic, molecular and condensate signatures in the atmospheres of exoplanets \citep[e.g.][]{VidalMadjar2003, Linsky2010, Fossati2010, VidalMadjar2013, Sing2015}. 

The WFC3 camera is one of the instruments currently onboard the HST, and it was installed in May 2009 during the 4$^\mathrm{th}$ servicing mission of HST. Since 2012, the WFC3 camera has been used in two different observing modes, the normal (staring) mode, where the telescope pointing is fixed on the target, and the spatial scanning mode, where the telescope is slewing during an exposure, causing the image or the spectrum of the target to move on the detector. The spatial scanning technique allows for a larger number of photons to be collected in a single exposure without the risk of saturation. As a result, overheads are reduced and the achieved signal-to-noise ratio (S/N) is increased. This observational strategy has already been successfully used to provide an increasing number of exoplanetary spectra \citep[e.g.][]{Deming2013, McCullough2014B2014ApJ...791...55M, Crouzet2014, Fraine2014, Knutson2014B2014Natur.505...66K, Knutson2014B2014ApJ...794..155K, Kreidberg2014B2014Natur.505...69K, Kreidberg2014B2014ApJ...793L..27K, Stevenson2014B2014Sci...346..838S, Kreidberg2015, Tsiaras2016B2016ApJ...820...99T, Line2016B2016AJ....152..203L, Wakeford2017B2017Sci...356..628W, Tsiaras2018, Wakeford2018, Wakeford2019}.

However, the scanning-mode spectroscopic images have a complicated structure and, therefore, their analysis towards extracting the transmission spectrum of a transiting planet, follows a large number of steps. Each one of these steps includes calibration data and assumptions, all of which could potentially affect the data in a different way, if they are not precise enough. The only way to evaluate the effectiveness of such processes is the use of calibrating sources, of known characteristics. In absence of such sources, simulated observations that include both the observed object and the systematics become very important. \wayne \footnote{\url{https://github.com/ucl-exoplanets/wayne}} \citep{Varley2017}, is a simulator for WFC3 spectroscopy which can reproduce most of the systematics seen in the WFC3 observations, but also mimic potential systematics that may or may not be present.

Here, we use \wayne\ to evaluate the effect of systematics on the final transmission spectrum of exoplanets. We test the effect of the well-known horizontal shifts \citep[e.g.][]{Deming2013}, in cases of imprecise estimates, but also the effect of two potential systematics that we do not know if they are present or not in our analysis. These are: a) spacecraft drifts during the spatial scanning and b) non-linearity effects beyond the correctable levels.

The datasets simulated are those of HD\,209458\,b (ID: 12181, PI: Drake Deming), first analysed by \cite{Deming2013}, and 55\,Cancri\,e (ID: 13665, PI: Bj{\"o}rn Benneke), first analysed by \cite{Tsiaras2016B2016ApJ...820...99T}. These two datasets have provided the most precise spectra so far currently in the literature. In addition, the HD\,209458\,b dataset shows very strong horizontal shifts (about one pixel), while the and 55\,Cancri\,e dataset is the longest spatial scan (about 350 pixels). For this reason we choose them as suitable laboratories for testing systematics.

The analysis of the simulated observations is carried out using our specialised software for the analysis of WFC3, spatially scanned spectroscopic images \citep{Tsiaras2016B2016ApJ...820...99T,  Tsiaras2016B2016ApJ...832..202T, Tsiaras2018}, which has been integrated into the \iraclis \footnote{\url{https://github.com/ucl-exoplanets/Iraclis}} package.
% INTRODUCTION

% THE WAYNE SIMULATOR
\section{GENERAL SET-UP} \label{sec:setup}

In all simulations, both the HD\,209458\,b and the 55\,Cancri\,e observations are simulated using the setup described in \cite{Varley2017}, including all the known systematics of WFC3. The input stellar spectrum is a PHOENIX model \citep{Allard2012, Baraffe2015} for a star as similar as possible to HD\,209458 ($T_\mathrm{eff} = 6065 \, \mathrm{K},  \mathrm{[Fe/H]} = 0.00 \, \mathrm{[dex]},  \mathrm{log}(g_*) = 4.361 \, \mathrm{[cgs]}$). The sky background, horizontal shifts, vertical shifts, exposure times and planetary spectra are configured to match those recovered from the real observations (for more details we refer the reader to \cite{Tsiaras2016B2016ApJ...832..202T} and \cite{Tsiaras2016B2016ApJ...820...99T} for details on HD\,209458\,b and on 55\,Cancri\,e, respectively).

For the sinusoidal scan speed variations, a period of 0.7\,seconds and an amplitude of 1.5 is used. In addition, the jitter noise level used is 0.02 pixels per second.
% THE WAYNE SIMULATOR

% HORIZONTAL SHIFTS
\newpage
\section{HORIZONTAL SHIFTS} \label{sec:xshifts}

\begin{figure}
	\centering
	\includegraphics[width=\columnwidth]{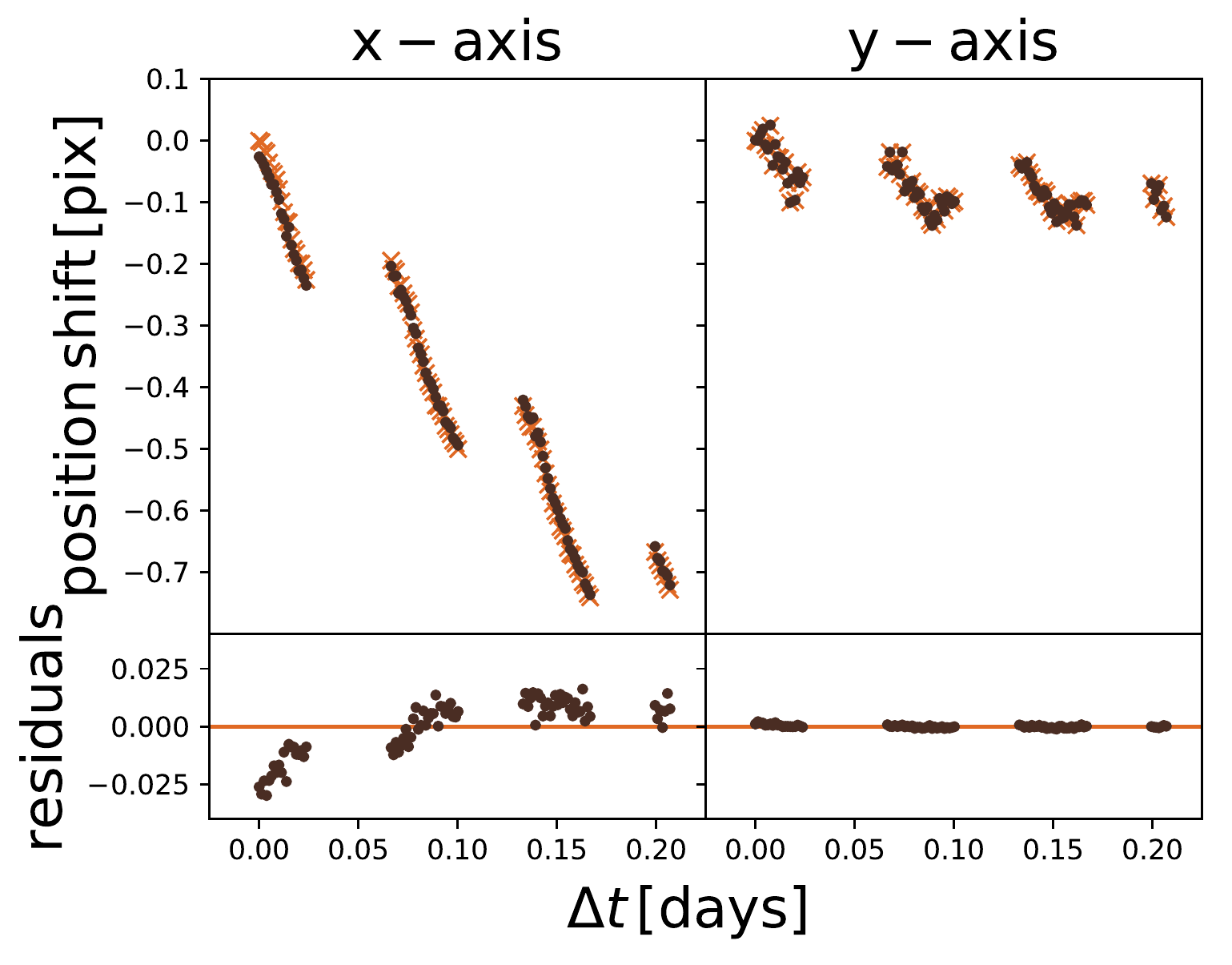}
	\caption{Disagreement between input and output horizontal position shifts. Despite being of the level of 0.025 pixels, such disagreement can lead to the misinterpretation of the molecular consistency of the planetary atmosphere.}
	\label{fig:xshifts_position}
\end{figure}

After a spatial scanned exposure, the HST guiding system fails to precisely reset the spectrum at its initial position. As a result, the observed spectrum is shifted horizontally, and the shift can be up to a few pixels over an entire visit \citep[e.g.][]{Deming2013, Kreidberg2014B2014Natur.505...69K, Kreidberg2014B2014ApJ...793L..27K, Fraine2014, Tsiaras2018}. Horizontal shifts are important as they displace the spectrum on the detector and also introduce additional systematics to the spectral light-curves, such as under-sampling \citep{Deming2013, Wilkins2014}. Vertical position shifts also exist \citep{Tsiaras2016B2016ApJ...832..202T, Tsiaras2016B2016ApJ...820...99T}, but here we focus only on the horizontal shifts.

Different calibration methods for the horizontal shifts exist in the literature \citep[e.g.][]{Deming2013, Kreidberg2014B2014ApJ...793L..27K, Haynes2015, Tsiaras2016B2016ApJ...832..202T, Evans2017}. The \wayne\ simulator offers a unique opportunity to test the efficiency of these calibration methods, case by case, by comparing the input, known, horizontal shifts with those calculated by a specific method. The efficiency of the calibration method used in \iraclis\ was tested in \cite{Varley2017} and found to be accurate within 0.5\% of the pixel. To calculate the horizontal shifts in \iraclis, we compare the structure of the first spatially scanned spectrum with all subsequent spectra, using the normalised sum along their columns, similarly to \cite{Kreidberg2014B2014ApJ...793L..27K}. For each consecutive image we interpolate and fit for the horizontal shift, relatively to the first one. The sums used above are corrected for the static (non wavelength-dependent) component of the flat-field, to avoid the bias introduced by its structure.

Here, we attempt to evaluate the bias in the final transmission spectrum, caused by ignoring the static (non wavelength-dependent) component of the flat-field, during the calibration for the horizontal shifts. It is beyond the scope of this work to validate other calibration methods used in the literature, but we will make the data sets used here available to the community, to enable cross-validation between different methods.

\begin{figure}
	\centering
	\includegraphics[width=\columnwidth]{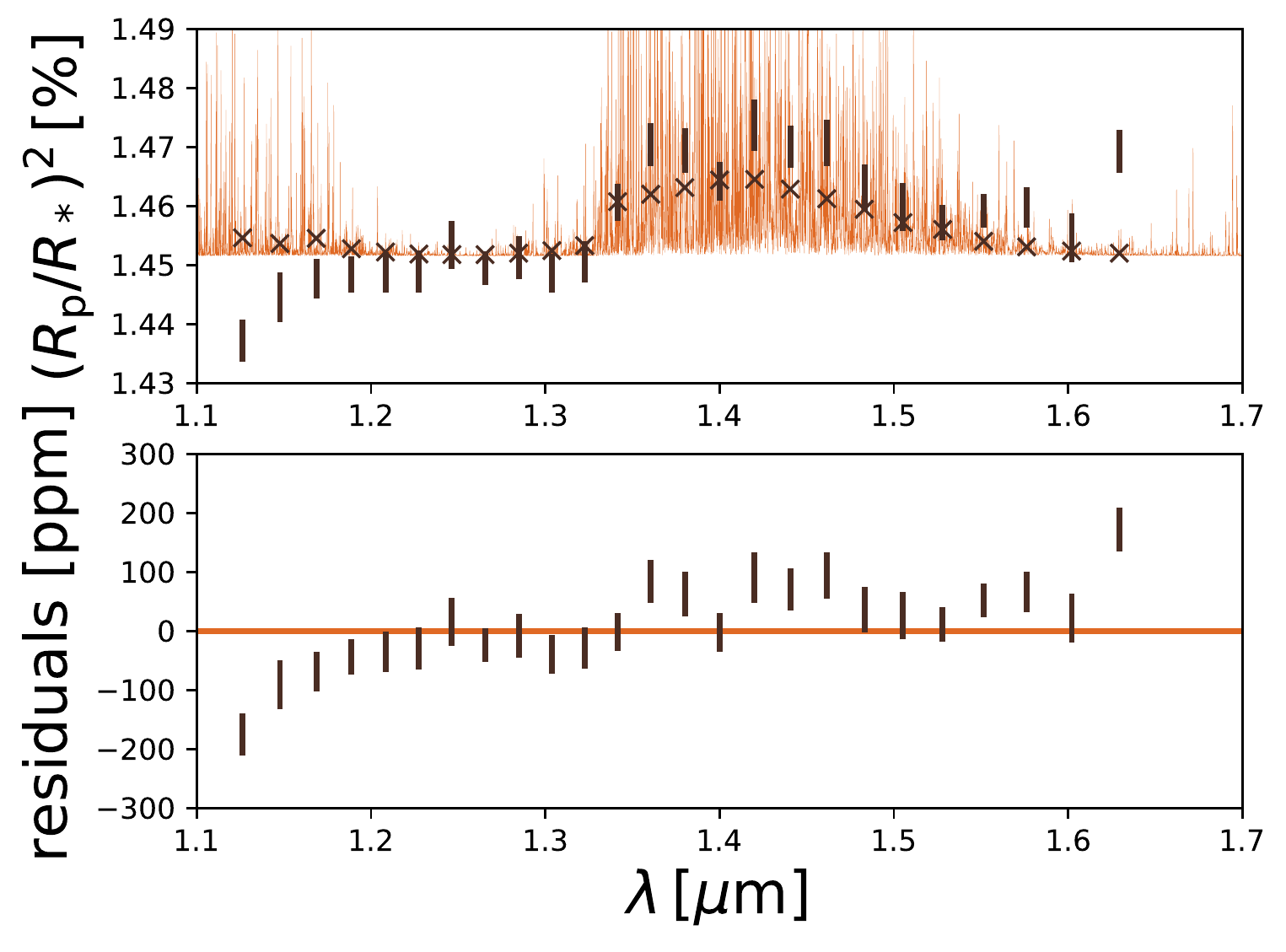}
	\caption{Comparison between input and output spectra where we can see the increasing trend towards longer wavelengths.}
	\label{fig:xshifts_spectrum}
\end{figure}

\subsection{Test and results}

For this test, we created a simulated version of the HD\,209458\,b observation with the general set-up described in Section \ref{sec:setup}. For the analysis, we modified \iraclis\ so that it ignores the non wavelength-dependent component of the flat-field.

\begin{figure}
	\centering
	\includegraphics[width=\columnwidth]{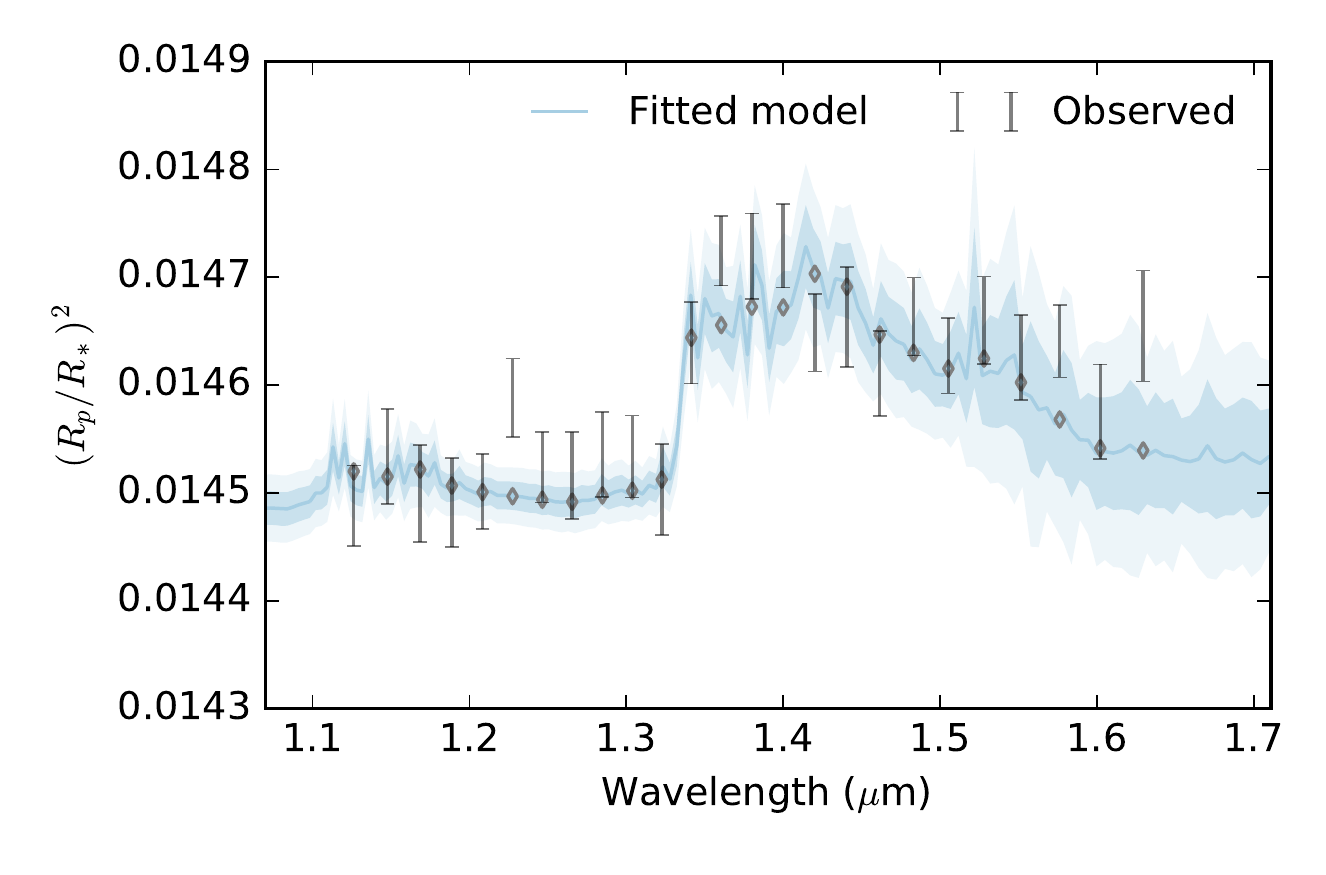}
	\includegraphics[width=1.1\columnwidth]{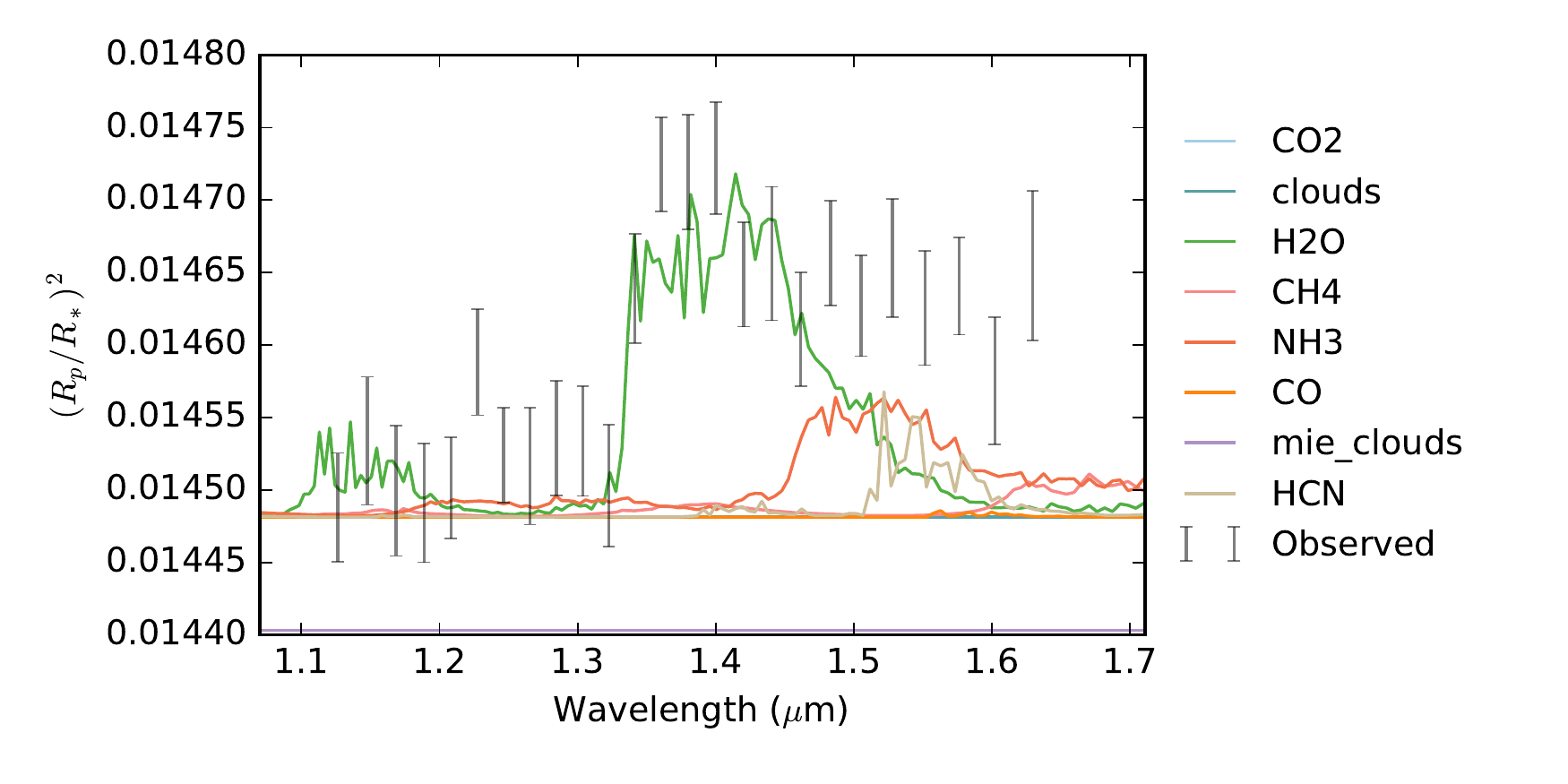}
	\caption{Top: Best-fit \taurex\ spectrum. Bottom: Contribution of the different molecules fitted on the HD\,209548\,b spectrum, where we can see the contribution from NH$_3$ and HCN.}
	\label{fig:xshifts_posteriors}
\end{figure}

Figure \ref{fig:xshifts_position} shows the discrepancies between the input, known, horizontal shifts with those calculated. As we can see here, the difference is non-uniform with time, starting from -2.5\% and increases to +1.0\% of a pixel. While this discrepancy appears to be small, we further tested the effect on the final spectrum and found that it causes an increasing trend with wavelength and a stronger water feature, Figure \ref{fig:xshifts_spectrum}. The final spectrum was analysed with \taurex\ \citep{Waldmann2015B2015ApJ...813...13W, Waldmann2015B2015ApJ...802..107W}, where two extra molecules where Identified: HCN and NH$_3$ (Figure \ref{fig:xshifts_posteriors}). These results indicate that even small biases in the horizontal shifts estimation can have an important impact on the final transmission spectrum, making the benchmarking of every calibration method used against simulations, necessary.
% HORIZONTAL SHIFTS

% IN-SCAN DRIFTS
\newpage
\section{IN-SCAN DRIFTS} \label{sec:inclined}

One assumption that is used for the wavelength calibration in \iraclis, is that during the spatial scanning, the star is moving parallel to the detectors columns \citep{Tsiaras2016B2016ApJ...832..202T}. Here, we attempt to evaluate the bias in the final transmission spectrum, caused if this assumption is not true.

\subsection{Test and results}

For this test, we created two simulated versions of the HD\,209458\,b observation and two simulated versions of the 55\,Cancri\,e observation with the general set-up described in Section \ref{sec:setup}, including a drift during the spatial scanning. More specifically, the datasets created had a drift of +2\% and -2\% (one for each planet). For the analysis, we used the standard version of \iraclis, without any modifications.

\begin{figure}
	\centering
	\includegraphics[width=\columnwidth]{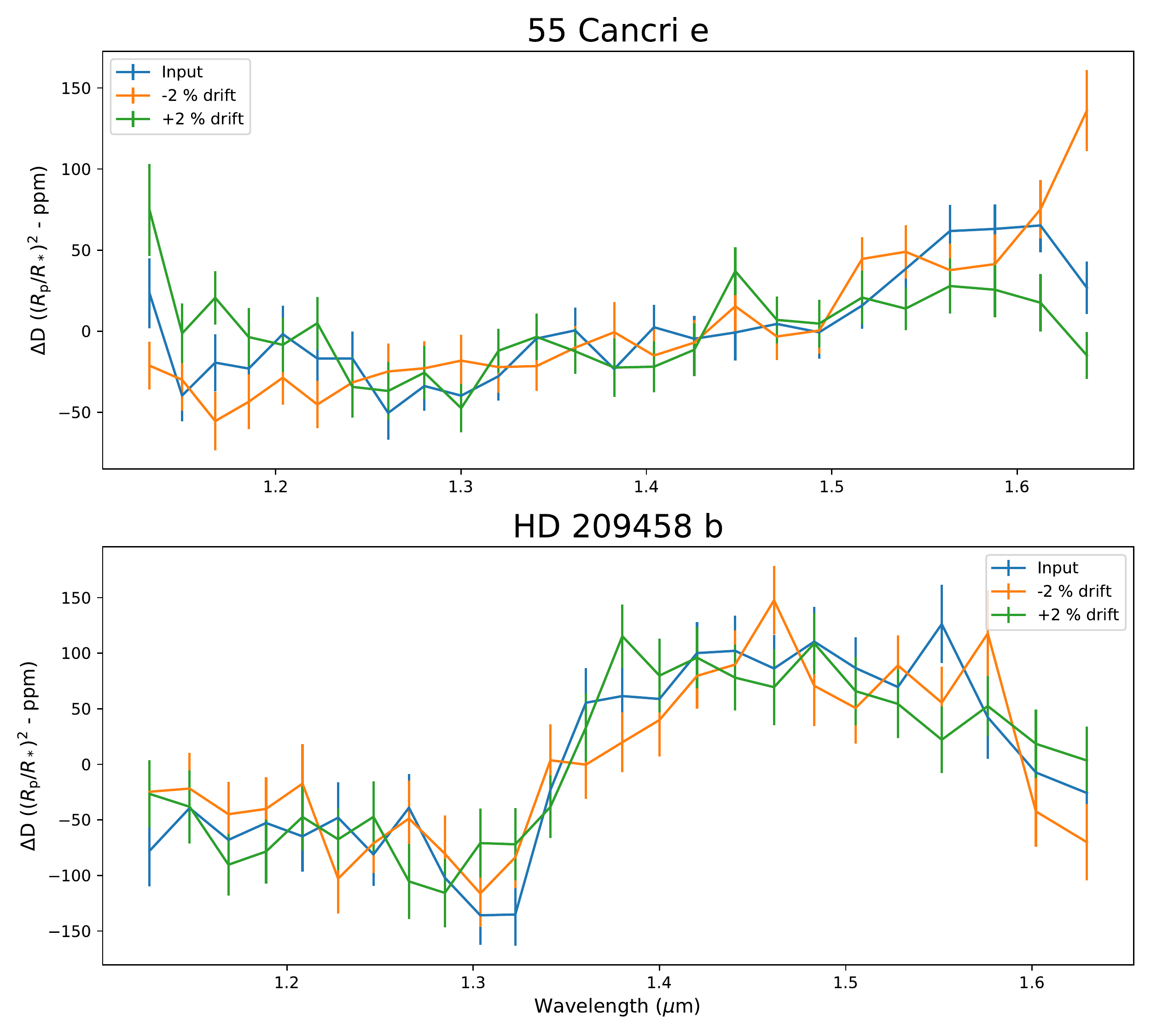}
	\caption{Differences between the input and output spectra for the different levels of in-scan drifts.}
	\label{fig:drift_spectra}
\end{figure}

Figure \ref{fig:drift_spectra} shows the discrepancies between the input, known, spectra and the output of the above test. As we can see here, the spectrum of HD\,209458\,b is not sensitive to the in-scan drifts, while the spectrum of 55\,Cancri\,e shows an increased slope for the negative drift and a reduced slope for the positive drift. This results shows that the very long spectra, like the one of 55\,Cancri\,e (350 pixels) are sensitive to in-scan drifts.

\subsection{Examining observations for in-scan drifts}

Given the above result, it is necessary to inspect very long spectra (above 300 pixels) for in-scan drifts. As it is not possible to locate the position of the target during the spatial scanning, we used simulated observations of \wayne\ to derive an observable feature that could be used to estimate the in-scan drift. We produced more simulated observations, with different levels of in-scan drifts, and concluded that the inclination of the blue-edge of the spectrum is linearly dependent to the in-scan drift, independently of the dataset. 

\begin{figure}
	\centering
	\includegraphics[width=0.9\columnwidth]{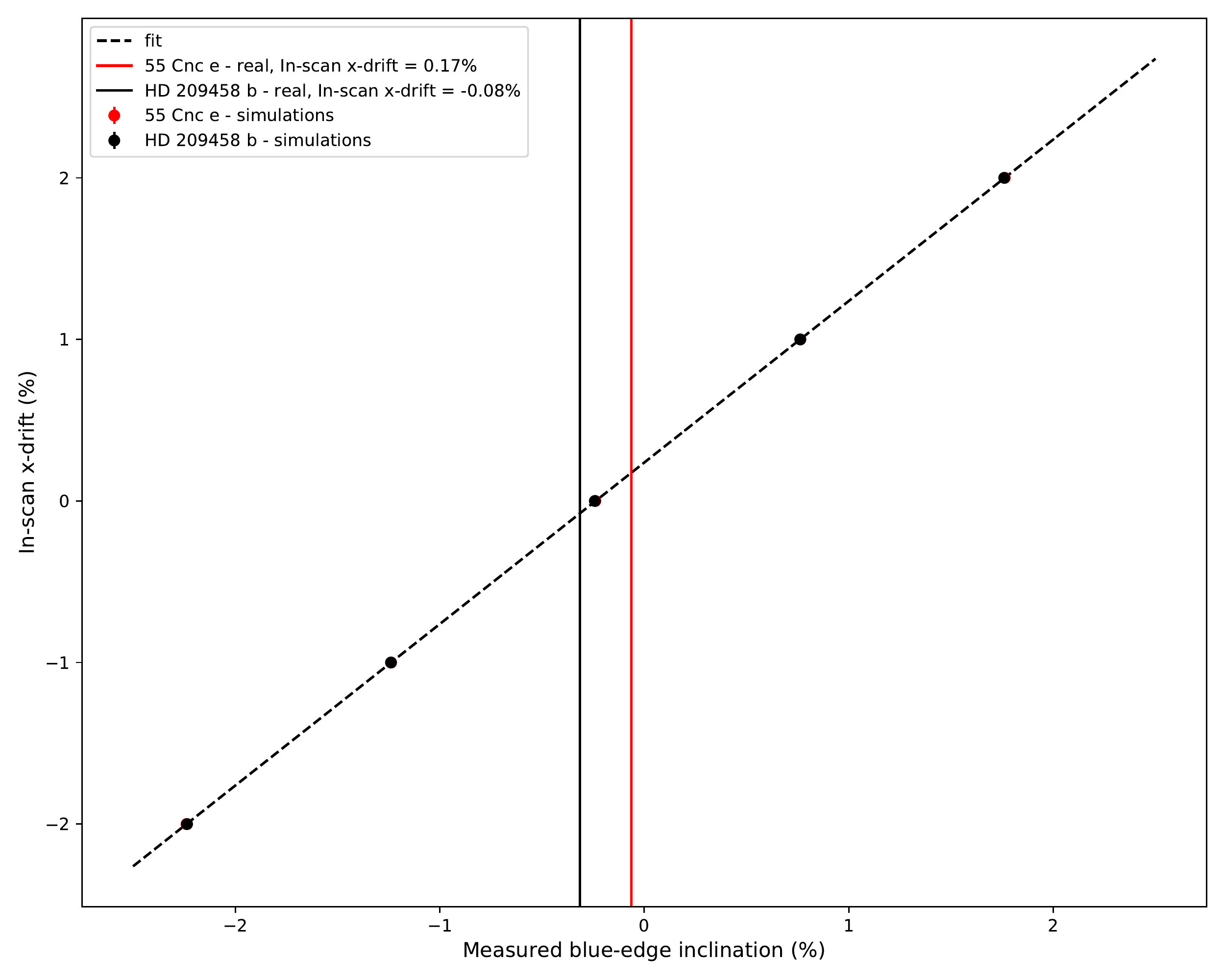}
	\caption{The linear relationship between the blue-edge inclination of the spectrum and the in-scan drifts.}
	\label{fig:drift_inclination}
\end{figure}

Figure \ref{fig:drift_inclination} shows the that the relationship between the two quantities is the same, both for the HD\,209458\,b datasets and for the 55\,Cancri\,e ones. From these values we estimated that the in-scan drift is -0.08\% for the real dataset of HD\,209458\,b, and 0.17\% for the real dataset of 55\,Cancri\,e. Also, from the simulated observations, it became clear that in-scan drifts up to a level of +/-1\% are to sufficient to affect the final spectra to a detectable level, hence, the current spectra of both planets are not affected by in-scan drifts.

% IN-SCAN DRIFTS

% NON-LINEARITY
\section{NON-LINEARITY} \label{sec:linearity}

The WFC3/IR detector, being an HgCdTe one does not perform linearly, meaning that the efficiency of the detector in recording incoming photons is decreasing with time, within an exposure. More specifically, the response of the detector is described, by a 4$^\mathrm{th}$ order polynomial:
\begin{equation}
	F_\mathrm{c} = (1+c_1 + c_2 F + c_3 F^2 + c_4 F^3) F
	\label{eq:linearity}
\end{equation}
\vspace{-0.3cm}

\noindent where $F$ are the recorded DNs (i.e. counts, or digital units), $F_\mathrm{c}$ is incoming DNs, and $c_{1-4}$ are the non-linearity coefficients \citep[][ \textit{u1k1727mi\_lin.fits} calibration file]{Hilbert2008B2008wfc..rept...39H}.

While the non-linear behaviour is pixel-dependent \citep{Hilbert2014}, only quadrant-based coefficients are available, meaning that there is a potential bias. According to \cite{Hilbert2008B2008wfc..rept...39H}, these coefficients can only correct the average non-linear behaviour of a pixel, across observations. However, on individual observations, this correction cannot reach the 0.3\% correction level which is ra requirement for WFC3. Here, we attempt to evaluate the bias in the final transmission spectrum, caused by potential differences in the non-linearity levels than those expected.

\subsection{Test and results}

For this test, we created two simulated versions of the HD\,209458\,b observation and two simulated versions of the 55\,Cancri\,e observation with the general set-up described in Section \ref{sec:setup}, including a variable non-linearity effect. More specifically, the datasets created had a non-linearity effect enhanced by 0.35\% and reduced by 0.35\% (one for each planet). To achieve that we modified the non-linearity coefficients $c_3$ and $c_4$. IN the reduced case, $c_3$ was multiplied by 0.9 and $c_4$ by 0.93232, while for the enhanced case, $c_3$ was multiplied by 1.1 and $c_4$ by 1.0612. The analysis, was, once more, carried out using the standard version of \iraclis, without any modifications.

\begin{figure}
	\centering
	\includegraphics[width=\columnwidth]{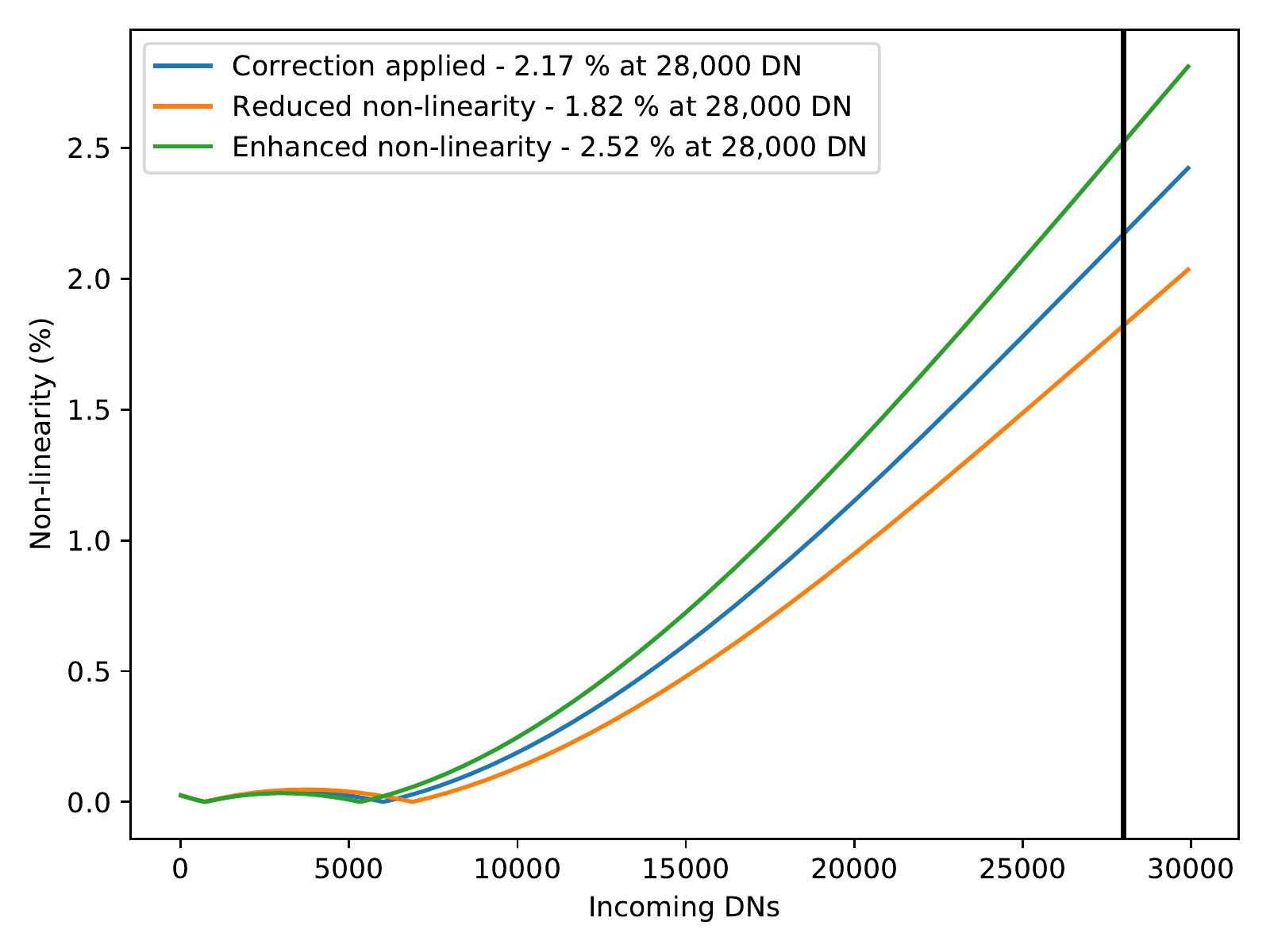}
	\caption{Non-linearity strength as a function of incoming flux. The blue curve indicates the correction applied while the other curves indicate the \wayne\ input in the two cases of enchanced and reduced non-linearity strength. The vertical line indicated the limit of 28,000 DNs -- i.e. 70,000 electrons.}
	\label{fig:linearity_test}
\end{figure}

\begin{figure}
	\centering
	\includegraphics[width=0.9\columnwidth]{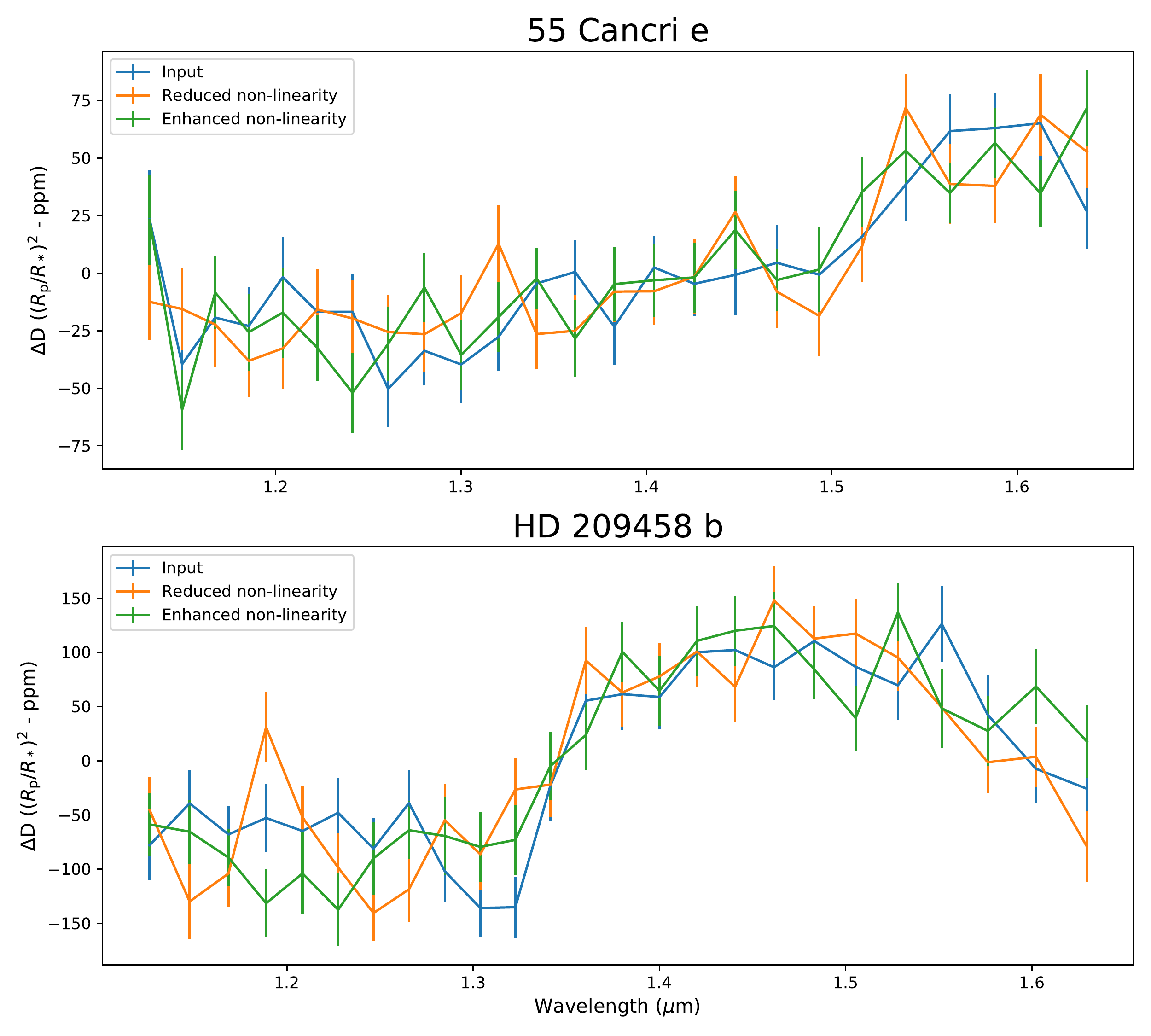}
	\caption{Differences between the input and output spectra for the different levels of non-linearity strengths.}
	\label{fig:linearity_spectra}
\end{figure}

Figure \ref{fig:linearity_test}  Figure \ref{fig:linearity_spectra} shows the discrepancies between the input, known, spectra and the output of the above test. As we can see here, the spectrum of HD\,209458\,b is not sensitive to the in-scan drifts, while the spectrum of 55\,Cancri\,e shows an increased slope for the negative drift and a reduced slope for the positive drift. This results shows that the very long spectra, like the one of 55\,Cancri\,e (350 pixels) are sensitive to in-scan drifts.

Figure \ref{fig:linearity_test} shows the difference in the input non-linearity effect and the expected (standard HST correction) fro the enhanced and the reduced cases. Figure \ref{fig:linearity_spectra} shows the discrepancies between the input, known, spectra and the output of the above test. As we can see here, both spectra are not sensitive to the different non-linearity strengths, ensuring that current non-linearity correction is sufficient for exoplanet observations, even in the cases where the total flux level is close to the highest acceptable limit of 28,000 DNs -- i.e. 70,000 electrons.
% NON-LINEARITY

% CONCLUSIONS
\section{CONCLUSIONS} \label{sec:conclusions}

In this work we tested the stability of current transmission spectra of exoplanets observed with the HST/WFC3 camera in the spacial scanning mode, and the effectiveness of data analysis techniques against three different kind of systematics, known and unknown. More specifically we test and found that:

\begin{itemize}
\item{horizontal shifts can strongly affect the final transmission spectra and should be calculated with very good precision (better that 1\% of a pixel),}
\item{in-scan drifts -- i.e. horizontal shifts during an exposure -- can affect long scans such as the 55\,Cancri\,e case but only if they exceed 1\%, for these reason future observations in this mode should be tested for  such drifts,}
\item{non-linearity variations, to the level of the known uncertainties do not affect our data, hence current corrections are sufficient.}
\end{itemize}

In addition, these tests demonstrate the robustness of the current spectra of HD\,209458\,b and 55\,Cancri\,e. In absence of calibrating sources, HST/WFC3 observations and data analysis techniques must be calibrated using simulated data, the only data for which we know the input exoplanetary signal. To serve their scope, these simulations need to be as close to the real data as possible, enabling us to  investigate known and unknown systematics. The \wayne\ simulator is such a tool, and here, we demonstrated how efficiently it can be used to test the efficiency of current pipelines and also rule out the possibility of our data being affected by certain systematics. 
% CONCLUSIONS

%AKNOLEDGEMENTS
\begin{acknowledgements}

This project has received funding from the European Research Council (ERC) under the European Union's Horizon 2020 research and innovation programme (grant agreements 758892, ExoAI) and under the European Union's Seventh Framework Programme (FP7/2007-2013)/ ERC grant agreement numbers 617119 (ExoLights). We furthermore acknowledge funding by the Science and Technology Funding Council (STFC) grants: ST/K502406/1 and ST/P000282/1.

\end{acknowledgements}
%AKNOLEDGEMENTS

%REFERENCES
{\small
\bibliographystyle{aasjournals}
\bibliography{references} 

\begin{thebibliography}{}
\expandafter\ifx\csname natexlab\endcsname\relax\def\natexlab#1{#1}\fi
\providecommand{\url}[1]{\href{#1}{#1}}
\providecommand{\dodoi}[1]{doi:~\href{http://doi.org/#1}{\nolinkurl{#1}}}
\providecommand{\doeprint}[1]{\href{http://ascl.net/#1}{\nolinkurl{http://ascl.net/#1}}}
\providecommand{\doarXiv}[1]{\href{https://arxiv.org/abs/#1}{\nolinkurl{https://arxiv.org/abs/#1}}}

\bibitem[{{Allard} {et~al.}(2012){Allard}, {Homeier}, \&
  {Freytag}}]{Allard2012}
{Allard}, F., {Homeier}, D., \& {Freytag}, B. 2012, Philosophical Transactions
  of the Royal Society of London Series A, 370, 2765,
  \dodoi{10.1098/rsta.2011.0269}

\bibitem[{{Baraffe} {et~al.}(2015){Baraffe}, {Homeier}, {Allard}, \&
  {Chabrier}}]{Baraffe2015}
{Baraffe}, I., {Homeier}, D., {Allard}, F., \& {Chabrier}, G. 2015, \aap, 577,
  A42, \dodoi{10.1051/0004-6361/201425481}

\bibitem[{{Charbonneau} {et~al.}(2002){Charbonneau}, {Brown}, {Noyes}, \&
  {Gilliland}}]{Charbonneau2002}
{Charbonneau}, D., {Brown}, T.~M., {Noyes}, R.~W., \& {Gilliland}, R.~L. 2002,
  \apj, 568, 377, \dodoi{10.1086/338770}

\bibitem[{{Crouzet} {et~al.}(2014){Crouzet}, {McCullough}, {Deming}, \&
  {Madhusudhan}}]{Crouzet2014}
{Crouzet}, N., {McCullough}, P.~R., {Deming}, D., \& {Madhusudhan}, N. 2014,
  \apj, 795, 166, \dodoi{10.1088/0004-637X/795/2/166}

\bibitem[{{Deming} {et~al.}(2013){Deming}, {Wilkins}, {McCullough}, {Burrows},
  {Fortney}, {Agol}, {Dobbs-Dixon}, {Madhusudhan}, {Crouzet}, {Desert},
  {Gilliland}, {Haynes}, {Knutson}, {Line}, {Magic}, {Mandell}, {Ranjan},
  {Charbonneau}, {Clampin}, {Seager}, \& {Showman}}]{Deming2013}
{Deming}, D., {Wilkins}, A., {McCullough}, P., {et~al.} 2013, \apj, 774, 95,
  \dodoi{10.1088/0004-637X/774/2/95}

\bibitem[{{Evans} {et~al.}(2017){Evans}, {Sing}, {Kataria}, {Goyal}, {Nikolov},
  {Wakeford}, {Deming}, {Marley}, {Amundsen}, {Ballester}, {Barstow},
  {Ben-Jaffel}, {Bourrier}, {Buchhave}, {Cohen}, {Ehrenreich}, {Garc{\'\i}a
  Mu{\~n}oz}, {Henry}, {Knutson}, {Lavvas}, {Lecavelier Des Etangs}, {Lewis},
  {L{\'o}pez-Morales}, {Mandell}, {Sanz-Forcada}, {Tremblin}, \&
  {Lupu}}]{Evans2017}
{Evans}, T.~M., {Sing}, D.~K., {Kataria}, T., {et~al.} 2017, \nat, 548, 58,
  \dodoi{10.1038/nature23266}

\bibitem[{{Fossati} {et~al.}(2010){Fossati}, {Haswell}, {Froning}, {Hebb},
  {Holmes}, {Kolb}, {Helling}, {Carter}, {Wheatley}, {Collier Cameron},
  {Loeillet}, {Pollacco}, {Street}, {Stempels}, {Simpson}, {Udry}, {Joshi},
  {West}, {Skillen}, \& {Wilson}}]{Fossati2010}
{Fossati}, L., {Haswell}, C.~A., {Froning}, C.~S., {et~al.} 2010, \apjl, 714,
  L222, \dodoi{10.1088/2041-8205/714/2/L222}

\bibitem[{{Fraine} {et~al.}(2014){Fraine}, {Deming}, {Benneke}, {Knutson},
  {Jord{\'a}n}, {Espinoza}, {Madhusudhan}, {Wilkins}, \&
  {Todorov}}]{Fraine2014}
{Fraine}, J., {Deming}, D., {Benneke}, B., {et~al.} 2014, \nat, 513, 526,
  \dodoi{10.1038/nature13785}

\bibitem[{{Haynes} {et~al.}(2015){Haynes}, {Mandell}, {Madhusudhan}, {Deming},
  \& {Knutson}}]{Haynes2015}
{Haynes}, K., {Mandell}, A.~M., {Madhusudhan}, N., {Deming}, D., \& {Knutson},
  H. 2015, \apj, 806, 146, \dodoi{10.1088/0004-637X/806/2/146}

\bibitem[{{Hilbert}(2008)}]{Hilbert2008B2008wfc..rept...39H}
{Hilbert}, B. 2008, {WFC3 TV3 Testing: IR Channel Nonlinearity Correction},
  Tech. rep.

\bibitem[{{Hilbert}(2014)}]{Hilbert2014}
---. 2014, {Updated non-linearity calibration method for WFC3/IR}, Tech. rep.

\bibitem[{{Knutson} {et~al.}(2014{\natexlab{a}}){Knutson}, {Benneke}, {Deming},
  \& {Homeier}}]{Knutson2014B2014Natur.505...66K}
{Knutson}, H.~A., {Benneke}, B., {Deming}, D., \& {Homeier}, D.
  2014{\natexlab{a}}, \nat, 505, 66, \dodoi{10.1038/nature12887}

\bibitem[{{Knutson} {et~al.}(2014{\natexlab{b}}){Knutson}, {Dragomir},
  {Kreidberg}, {Kempton}, {McCullough}, {Fortney}, {Bean}, {Gillon}, {Homeier},
  \& {Howard}}]{Knutson2014B2014ApJ...794..155K}
{Knutson}, H.~A., {Dragomir}, D., {Kreidberg}, L., {et~al.} 2014{\natexlab{b}},
  \apj, 794, 155, \dodoi{10.1088/0004-637X/794/2/155}

\bibitem[{{Kreidberg} {et~al.}(2014{\natexlab{a}}){Kreidberg}, {Bean},
  {D{\'e}sert}, {Benneke}, {Deming}, {Stevenson}, {Seager}, {Berta-Thompson},
  {Seifahrt}, \& {Homeier}}]{Kreidberg2014B2014Natur.505...69K}
{Kreidberg}, L., {Bean}, J.~L., {D{\'e}sert}, J.-M., {et~al.}
  2014{\natexlab{a}}, \nat, 505, 69, \dodoi{10.1038/nature12888}

\bibitem[{{Kreidberg} {et~al.}(2014{\natexlab{b}}){Kreidberg}, {Bean},
  {D{\'e}sert}, {Line}, {Fortney}, {Madhusudhan}, {Stevenson}, {Showman},
  {Charbonneau}, {McCullough}, {Seager}, {Burrows}, {Henry}, {Williamson},
  {Kataria}, \& {Homeier}}]{Kreidberg2014B2014ApJ...793L..27K}
---. 2014{\natexlab{b}}, \apjl, 793, L27, \dodoi{10.1088/2041-8205/793/2/L27}

\bibitem[{{Kreidberg} {et~al.}(2015){Kreidberg}, {Line}, {Bean}, {Stevenson},
  {D{\'e}sert}, {Madhusudhan}, {Fortney}, {Barstow}, {Henry}, {Williamson}, \&
  {Showman}}]{Kreidberg2015}
{Kreidberg}, L., {Line}, M.~R., {Bean}, J.~L., {et~al.} 2015, \apj, 814, 66,
  \dodoi{10.1088/0004-637X/814/1/66}

\bibitem[{{Line} {et~al.}(2016){Line}, {Stevenson}, {Bean}, {Desert},
  {Fortney}, {Kreidberg}, {Madhusudhan}, {Showman}, \&
  {Diamond-Lowe}}]{Line2016B2016AJ....152..203L}
{Line}, M.~R., {Stevenson}, K.~B., {Bean}, J., {et~al.} 2016, \aj, 152, 203,
  \dodoi{10.3847/0004-6256/152/6/203}

\bibitem[{{Linsky} {et~al.}(2010){Linsky}, {Yang}, {France}, {Froning},
  {Green}, {Stocke}, \& {Osterman}}]{Linsky2010}
{Linsky}, J.~L., {Yang}, H., {France}, K., {et~al.} 2010, \apj, 717, 1291,
  \dodoi{10.1088/0004-637X/717/2/1291}

\bibitem[{{McCullough} {et~al.}(2014){McCullough}, {Crouzet}, {Deming}, \&
  {Madhusudhan}}]{McCullough2014B2014ApJ...791...55M}
{McCullough}, P.~R., {Crouzet}, N., {Deming}, D., \& {Madhusudhan}, N. 2014,
  \apj, 791, 55, \dodoi{10.1088/0004-637X/791/1/55}

\bibitem[{{Sing} {et~al.}(2015){Sing}, {Wakeford}, {Showman}, {Nikolov},
  {Fortney}, {Burrows}, {Ballester}, {Deming}, {Aigrain}, {D{\'e}sert},
  {Gibson}, {Henry}, {Knutson}, {Lecavelier des Etangs}, {Pont},
  {Vidal-Madjar}, {Williamson}, \& {Wilson}}]{Sing2015}
{Sing}, D.~K., {Wakeford}, H.~R., {Showman}, A.~P., {et~al.} 2015, \mnras, 446,
  2428, \dodoi{10.1093/mnras/stu2279}

\bibitem[{{Stevenson} {et~al.}(2014){Stevenson}, {D{\'e}sert}, {Line}, {Bean},
  {Fortney}, {Showman}, {Kataria}, {Kreidberg}, {McCullough}, {Henry},
  {Charbonneau}, {Burrows}, {Seager}, {Madhusudhan}, {Williamson}, \&
  {Homeier}}]{Stevenson2014B2014Sci...346..838S}
{Stevenson}, K.~B., {D{\'e}sert}, J.-M., {Line}, M.~R., {et~al.} 2014, Science,
  346, 838, \dodoi{10.1126/science.1256758}

\bibitem[{{Tsiaras} {et~al.}(2016{\natexlab{a}}){Tsiaras}, {Waldmann},
  {Rocchetto}, {Varley}, {Morello}, {Damiano}, \&
  {Tinetti}}]{Tsiaras2016B2016ApJ...832..202T}
{Tsiaras}, A., {Waldmann}, I.~P., {Rocchetto}, M., {et~al.} 2016{\natexlab{a}},
  \apj, 832, 202, \dodoi{10.3847/0004-637X/832/2/202}

\bibitem[{{Tsiaras} {et~al.}(2016{\natexlab{b}}){Tsiaras}, {Rocchetto},
  {Waldmann}, {Venot}, {Varley}, {Morello}, {Damiano}, {Tinetti}, {Barton},
  {Yurchenko}, \& {Tennyson}}]{Tsiaras2016B2016ApJ...820...99T}
{Tsiaras}, A., {Rocchetto}, M., {Waldmann}, I.~P., {et~al.} 2016{\natexlab{b}},
  \apj, 820, 99, \dodoi{10.3847/0004-637X/820/2/99}

\bibitem[{{Tsiaras} {et~al.}(2018){Tsiaras}, {Waldmann}, {Zingales},
  {Rocchetto}, {Morello}, {Damiano}, {Karpouzas}, {Tinetti}, {McKemmish},
  {Tennyson}, \& {Yurchenko}}]{Tsiaras2018}
{Tsiaras}, A., {Waldmann}, I.~P., {Zingales}, T., {et~al.} 2018, \aj, 155,
  \dodoi{10.3847/1538-3881/aaaf75}

\bibitem[{{Varley} {et~al.}(2017){Varley}, {Tsiaras}, \&
  {Karpouzas}}]{Varley2017}
{Varley}, R., {Tsiaras}, A., \& {Karpouzas}, K. 2017, \apjs, 231, 13,
  \dodoi{10.3847/1538-4365/aa7750}

\bibitem[{{Vidal-Madjar} {et~al.}(2003){Vidal-Madjar}, {Lecavelier des Etangs},
  {D{\'e}sert}, {Ballester}, {Ferlet}, {H{\'e}brard}, \&
  {Mayor}}]{VidalMadjar2003}
{Vidal-Madjar}, A., {Lecavelier des Etangs}, A., {D{\'e}sert}, J.-M., {et~al.}
  2003, \nat, 422, 143, \dodoi{10.1038/nature01448}

\bibitem[{{Vidal-Madjar} {et~al.}(2013){Vidal-Madjar}, {Huitson}, {Bourrier},
  {D{\'e}sert}, {Ballester}, {Lecavelier des Etangs}, {Sing}, {Ehrenreich},
  {Ferlet}, {H{\'e}brard}, \& {McConnell}}]{VidalMadjar2013}
{Vidal-Madjar}, A., {Huitson}, C.~M., {Bourrier}, V., {et~al.} 2013, \aap, 560,
  A54, \dodoi{10.1051/0004-6361/201322234}

\bibitem[{{Wakeford} {et~al.}(2017){Wakeford}, {Sing}, {Kataria}, {Deming},
  {Nikolov}, {Lopez}, {Tremblin}, {Amundsen}, {Lewis}, {Mandell}, {Fortney},
  {Knutson}, {Benneke}, \& {Evans}}]{Wakeford2017B2017Sci...356..628W}
{Wakeford}, H.~R., {Sing}, D.~K., {Kataria}, T., {et~al.} 2017, Science, 356,
  628, \dodoi{10.1126/science.aah4668}

\bibitem[{{Wakeford} {et~al.}(2018){Wakeford}, {Sing}, {Deming}, {Lewis},
  {Goyal}, {Wilson}, {Barstow}, {Kataria}, {Drummond}, {Evans}, {Carter},
  {Nikolov}, {Knutson}, {Ballester}, \& {Mand ell}}]{Wakeford2018}
{Wakeford}, H.~R., {Sing}, D.~K., {Deming}, D., {et~al.} 2018, \aj, 155, 29,
  \dodoi{10.3847/1538-3881/aa9e4e}

\bibitem[{{Wakeford} {et~al.}(2019){Wakeford}, {Lewis}, {Fowler}, {Bruno},
  {Wilson}, {Moran}, {Valenti}, {Batalha}, {Filippazzo}, {Bourrier},
  {H{\"o}rst}, {Lederer}, \& {de Wit}}]{Wakeford2019}
{Wakeford}, H.~R., {Lewis}, N.~K., {Fowler}, J., {et~al.} 2019, \aj, 157, 11,
  \dodoi{10.3847/1538-3881/aaf04d}

\bibitem[{{Waldmann} {et~al.}(2015{\natexlab{a}}){Waldmann}, {Rocchetto},
  {Tinetti}, {Barton}, {Yurchenko}, \&
  {Tennyson}}]{Waldmann2015B2015ApJ...813...13W}
{Waldmann}, I.~P., {Rocchetto}, M., {Tinetti}, G., {et~al.} 2015{\natexlab{a}},
  \apj, 813, 13, \dodoi{10.1088/0004-637X/813/1/13}

\bibitem[{{Waldmann} {et~al.}(2015{\natexlab{b}}){Waldmann}, {Tinetti},
  {Rocchetto}, {Barton}, {Yurchenko}, \&
  {Tennyson}}]{Waldmann2015B2015ApJ...802..107W}
{Waldmann}, I.~P., {Tinetti}, G., {Rocchetto}, M., {et~al.} 2015{\natexlab{b}},
  \apj, 802, 107, \dodoi{10.1088/0004-637X/802/2/107}

\bibitem[{{Wilkins} {et~al.}(2014){Wilkins}, {Deming}, {Madhusudhan},
  {Burrows}, {Knutson}, {McCullough}, \& {Ranjan}}]{Wilkins2014}
{Wilkins}, A.~N., {Deming}, D., {Madhusudhan}, N., {et~al.} 2014, \apj, 783,
  113, \dodoi{10.1088/0004-637X/783/2/113}

\end{thebibliography}
}
%REFERENCES

\end{document}